\documentclass{PoS}

\let\l\left
\let\r\right
\let\de\partial
\let\mrm\mathrm

\let\mbb\mathbb

\let\dag\dagger

\newcommand{\beq}{\begin{equation}}
\newcommand{\eeq}{\end{equation}}
\newcommand{\beqa}{\begin{eqnarray}}
\newcommand{\eeqa}{\end{eqnarray}}
\newcommand{\ba}{\begin{array}}
\newcommand{\ea}{\end{array}}
\newcommand{\bmat}{\begin{pmatrix}}
\newcommand{\emat}{\end{pmatrix}}
\newcommand{\bcas}{\begin{cases}}
\newcommand{\ecas}{\end{cases}}

\newcommand{\muh}{{\hat{\mu}}}
\newcommand{\nuh}{{\hat{\nu}}}
\newcommand{\rhoh}{{\hat{\rho}}}

\title{Thimble regularization at work for Gauge Theories: from toy models onwards.}

\ShortTitle{Thimble regularization at work for Gauge Theories: from toy models onwards.}

\author{\speaker{Francesco Di Renzo}\\
        University of Parma and INFN\\
        E-mail: \email{francesco.direnzo@unipr.it}}

\author{Giovanni Eruzzi\\
        Universit\`a di Parma and INFN \\
        E-mail: \email{giovanni.eruzzi@fis.unipr.it}}

\abstract{A final goal for thimble regularization of lattice field
theories is the application to lattice QCD and the study
of its phase diagram.
Gauge theories pose a number of conceptual and algorithmic
problems, some of which can be addressed even in the
framework of toy models. We report on our progresses in
this field, starting in particular from first successes in
the study of one link models.}

\FullConference{The 33rd International Symposium on Lattice Field Theory\\
		 14 -18 July  2015\\
		 Kobe International Conference Center, Kobe, Japan}

\begin{document}

\section{Thimble regularization in a nutshell}
\label{sec:basics}

Thimble regularization has been proposed in the broad context of extending
our capabilities to properly define quantum field theories
\cite{Witten}. It has been later applied to lattice field theories as
an attempt to overcome the sign problem \cite{AUthimble, Kiku}. In a
nutshell, it relies on the complexification of the original
degrees of freedom. The functional integral is defined on manifolds which
roughly speaking emerge as the generalization of Steepest Descent (SD)
paths. Not surprisingly, such an approach displays the same virtues of
saddle point evaluation of integrals, i.e. stationary phase and
localization of important contributions.

Morse theory \cite{MorseTHEORY} states that under suitable 
(but not so strict) conditions 
on holomorphic functions $S(x) = S_R + iS_I$ and $O(x)$
\begin{equation} \label{eq:Morse}
\int_{\cal C} dx \; O(x) \; \mbox{e}^{-S(x)} = \sum_\sigma n_\sigma \,
\mbox{e}^{-iS_I(p_\sigma)} \int_{\mathcal{J}_\sigma} dz \; O(z) \; \mbox{e}^{-S_R(z)}
\end{equation}
where the notation ($S,O$) alludes to a functional integral (even if 
there is no normalizing factor $Z^{-1}$). 
The content of (\ref{eq:Morse}) has to be understood as follows:
\begin{itemize}
\item the greek index $\sigma$ counts the stationary points
  $p_\sigma$ of the complex function $S(z)$;
\item each (stable) thimble $\mathcal{J}_\sigma$ is the union of 
all the Steepest Ascent (SA) paths falling
  into $p_\sigma$ at (minus) infinite time, {\em i.e.} the union of the
  solutions of 
\begin{eqnarray}\label{eq:thimblEQ}
\frac{dx_i}{d\tau} &=& \frac{\partial S_R(x,y)}{\partial x_i} \nonumber \\
\frac{dy_i}{d\tau} &=& \frac{\partial S_R(x,y)}{\partial y_i} 
\end{eqnarray}
satisfying $z(\tau=-\infty) = x(\tau=-\infty) + iy(\tau=-\infty) =
p_\sigma$;
\item in the homological sense 
${\cal C}=\sum_\sigma n_\sigma \mathcal{J}_\sigma$ and the thimbles
have the same real dimension of the original domain of integration;
\item the coefficients $n_\sigma$ count the intersections of the 
  unstable thimbles $\mathcal{K}_\sigma$ with the original domain 
  of integration; unstable thimbles are solutions of
  (\ref{eq:thimblEQ}) with $z(\tau=\infty) = p_\sigma$.
\item the imaginary part $S_I$
stays constant on a thimble; on the other side $S_R$ increases along
the SA solutions of (\ref{eq:thimblEQ}), thus ensuring convergence 
of (\ref{eq:Morse}). 
\end{itemize}

For the following it is useful to remind the reader of a
constructive approach. Near a critical point a field configuration can
be expressed as $\Phi_i = \phi_i-\phi_{\sigma,i}$; the real part of
the action is in turn 
$$
S_R\left(\phi\right)=S_R\left(\phi_\sigma\right)+\frac{1}{2}\Phi^T H \Phi+\mathcal{O}\left(\phi^3\right) 
$$
where $H$ is the hessian 
$$
H_{ij}=\frac{\delta^2S_R}{\delta\phi_i\delta\phi_j}\biggl|_{\phi=\phi_\sigma}
$$
which can be put in diagonal form 
$$
H=W\Lambda W^T
\;\;\;\;\;
\Lambda=\mathrm{diag}\left(\lambda_1,\cdots,\lambda_n,-\lambda_1,\cdots,-\lambda_n\right)
$$
once its eigenvalues are known and arranged in the matrix $W$. The
reader should notice the form of the spectrum: there is an equal
number of positive and negative eigenvalues. Half of the eigenvectors 
of the Hessian (those corresponding to positive eigenvalues) span 
the tangent space at the stable thimble at the critical point: if one leaves 
the critical point along those directions integrating the SA
equations, one covers the stable thimble. On the other side, the other 
directions take you along the unstable thimble.

At a generic point we lack an {\em a priori} knowledge of the tangent
space. The (local) basis can be nevertheless obtained by transporting
along the flow the basis at the critical point (see
\cite{AUthimble} or \cite{Kiku} for details). By doing this one also
realizes that the relative orientation between the canonical complex 
volume form and the real volume form, characterizing the tangent space 
of the thimble, contributes a phase to the integral. This is termed
the {\em residual phase} (see \cite{resPHASE} for details). 

\section{Thimble regularization for gauge theories}

\subsection{The basic set-up}
We now want to sketch the thimble construction for $\mathrm{SU}(N)$ gauge
theories defined by an action $S[U]$. Going to complex fields means
$$
\mrm{SU}\l(N\r)\ni U=e^{i x_aT^a}\rightarrow e^{i z_aT^a}=e^{i \l(x_a+i y_a\r)T^a}\in\mrm{SL}\l(N,\mbb{C}\r)
$$
with the caveat that 
$$
\mathrm{SU}\left(N\right)\ni U^\dag=e^{-i x_aT^a}\rightarrow e^{-i z_aT^a}=e^{-i \left(x_a+i y_a\right)T^a}=U^{-1}\in\mathrm{SL}\left(N,\mathbb{C}\right).
$$
Main ingredient is the Lie derivative
$$
\nabla^af\left(U\right)=\lim_{\alpha\rightarrow 0}\frac{1}{\alpha}\left[f\left(e^{i\alpha T^a}U\right)-f\left(U\right)\right]=\frac{\delta}{\delta\alpha}f\left(e^{i\alpha T^a}U\right)\biggl|_{\alpha=0}
$$
in terms of which we can write the SA equations as 
\beq \label{eq:SAgauge}
\frac{\mrm{d}}{\mrm{d}\tau}U_\muh\l(n;\tau\r)=\l(i\,T^a\bar{\nabla}_{n,\muh}^a\overline{S\l[U\l(\tau\r)\r]}\r)U_\muh\l(n;\tau\r).
\eeq
The solutions of these equations display the main properties we
expect. Namely, since $\frac{\mrm{d}}{\mrm{d}\tau}=
\bar{\nabla}_{n,\muh}^a\bar{S}\,\nabla_{n,\muh}^a+
\nabla_{n,\muh}^aS\,\bar{\nabla}_{n,\muh}^a$ we have that 
$$
\frac{\mrm{d}S^R}{\mrm{d}\tau}=\frac{1}{2}\frac{\mrm{d}}{\mrm{d}\tau}\l(S+\bar{S}\r)=\frac{1}{2}\l(\bar{\nabla}_{n,\muh}^a\bar{S}\,\nabla_{n,\muh}^a S+\nabla_{n,\muh}^aS\,\bar{\nabla}_{n,\muh}^a\bar{S}\r)=\left\Vert\nabla S\right\Vert^2\geq0
$$
and
$$
\frac{\mrm{d}S^I}{\mrm{d}\tau}=\frac{1}{2i}\frac{\mrm{d}}{\mrm{d}\tau}\l(S-\bar{S}\r)=\frac{1}{2i}\l(\bar{\nabla}_{n,\muh}^a\bar{S}\,\nabla_{n,\muh}^a S-\nabla_{n,\muh}^aS\,\bar{\nabla}_{n,\muh}^a\bar{S}\r)=0.
$$
Lie derivatives obey non-trivial commutation relations
\beqa
\nonumber\l[\nabla_{n,\muh}^a\,,\nabla_{m,\nuh}^b\r] &=& -f^{abc}\,\nabla_{n,\muh}^c\,\delta_{n,m}\delta_{\muh,\nuh}\\
\nonumber\l[\bar{\nabla}_{n,\muh}^a\,,\bar{\nabla}_{m,\nuh}^b\r] &=& -f^{abc}\,\bar{\nabla}_{n,\muh}^c\,\delta_{n,m}\delta_{\muh,\nuh}\\
\nonumber\l[\nabla_{n,\muh}^a\,,\bar{\nabla}_{m,\nuh}^b\r] &=& 0
\eeqa
from which we can get commutation relations for vectors 
$V\equiv
V_{n,\muh,a}\nabla_{n,\muh}^a+\bar{V}_{n,\muh,a}\bar{\nabla}_{n,\muh}^a$
$$
\l[V,V'\r]_{n,\muh,c}=-f^{abc}\,V_{n,\muh,a}V'_{n,\muh,b}.
$$
Taking $V'_{n,\muh,c}=\bar{\nabla}_{n,\muh}^c\bar{S}$ we can derive
the equation for transporting a vector $V$ from the
critical point to any point along the flow described by (\ref{eq:SAgauge})
\beq
\frac{\mrm{d}}{\mrm{d}\tau}V_{n,\muh,c}=\bar{\nabla}_{m,\nuh}^a\bar{\nabla}_{n,\muh}^c\bar{S}\,\bar{V}_{m,\nuh,a}+f^{abc}\,\bar{\nabla}_{n,\muh}^b\bar{S}\,V_{n,\muh,a}.
\eeq
Apparently we collected all the tools needed for the constructive
approach to thimbles we described in the previous section. 
In particular, the last equation we wrote would enable us 
to transport along the flow the basis of the tangent space at 
the critical point. As a matter of fact all this is still void, 
because we still miss a proper definition for thimbles. 
While till now everything seems to be quite natural, we 
soon realize we need new ingredients to generalize the construction of
thimbles in the case of gauge theories.

\subsection{From non-degenerate critical points to non-degenerate critical submanifolds}

Once local gauge invariance is in place, every stationary point of a
gauge-invariant action belongs to a manifold of stationary points and, 
in particular, the Hessian is degenerate. The relevant picture is now 
provided \cite{AtBott} by generalizing the concept of a
non-degenerate critical point\footnote{The stationary points of 
Section \ref{sec:basics} were non-degenerate critical points: the 
hessian had no zero eigenvalue.} into that of a 
{\em non-degenerate critical manifold} \cite{AtBott}.  
A manifold ${\cal N} \subset {\cal C}$ is a non-degenerate critical sub-manifold
of 
${\cal C}$ for the function $F: \; {\cal C}\rightarrow \mathbb{R}$ if:
\begin{itemize}
\item[1.] $d F = 0$ along  ${\cal N}$;
\item[2.] The Hessian $\partial^2 F$ is non-degenerate on the normal bundle $\nu({\cal N})$.
\end{itemize}
If we consider the $A=0$ vacuum of an $SU(N_c)$ Yang-Mills theory ($F$
being given by the action $S$), ${\cal N}^{(0)}$ is given by the complete gauge
orbit associated to $A=0$, the real dimension of such critical
sub-manifold being given by $(V-1)(N_c^2-1)$. 

When we complexify, we switch to $SL(N_c,\mbb{C})$ and we get instead
the gauge orbit in the latter. As for non-zero eigenvalues of the
action hessian, we get an equal number of positive and negative
eigenvalues. These are once again associated to the SA and SD flows 
described by (\ref{eq:SAgauge}) (and we will be once again left with the
right real dimension of the thimbles).

All in all, the thimble {\em e.g.} associated to $A=0$ for the
$SU(3)$ Yang-Mills action is defined by
$$
{\cal J}_0:= 
\left\{ 
U\in (SL(3,\mathbb{C}))^{4V} \; | \; \; 
\exists U(\tau) \; \; \mbox{ solution of Eq.~(\ref{eq:SAgauge})}
\; \; | \; \; 
U(0)=U 
\; \; \& \; \; 
\lim_{\tau\rightarrow - \infty} U(\tau) \in {\cal N}^{(0)} 
\right\}.
$$
The meaning of the construction we sketched gets clearer when one
realizes that under $SL(3,\mbb{C})$ gauge transformations
$U_{\nu}(x) \rightarrow \Lambda(x) U_{\nu}(x;\tau) \Lambda(x+\hat{\nu})^{-1}$
$$
(T_a \overline{\nabla}_{x,\nu,a} \overline{S[U]}) \rightarrow 
\left(\Lambda(x)^{-1}\right)^{\dag} 
(T_a \overline{\nabla}_{x,\nu,a} \overline{S[U]}) 
\Lambda(x)^{\dag}.
$$
This means that we have the SA covariant only provided
$\Lambda(x)^\dagger=\Lambda(x)^{-1}$, 
{\em i.e.} for $SU(3)$ transformations. 
This also means that if we take a SA from $A=0$, at any stage we can 
perform a gauge transformation and this will take us to a point
starting from which under SD we are going to eventually 
land on another point on the gauge orbit of $A=0$
(decided by the gauge transformation we choose). For further 
details on the construction we sketched the reader is 
referred to \cite{AUthimble}.

\subsection{Pure Yang Mills SU(N): torons and all that}

Thimble regularization is {\em per se} an interesting
subject. Nevertheless, the main motivation for probing it as a
solution of the sign problem is to eventually tackle the investigation
of the QCD phase diagram. It is thus of outmost importance that the
framework we have just sketched is proven to be effective for gauge
theories. Before we mention the results we have already got for simple
toy models, we now discuss a possible application that displays 
a few of the subtleties one should be ready to face in the case of
gauge theories.

A somehow artificial sign problem can be encountered by addressing the study
of the Wilson action
$$
S_G\l[U\r]=\beta\sum_{m\in\Lambda}\sum_{\rhoh<\nuh}\l[1-\frac{1}{2N}\mrm{Tr}\l(U_{\rhoh\nuh}\l(m\r)+U_{\rhoh\nuh}^{-1}\l(m\r)\r)\r]
$$
at complex values of the coupling $\beta$. Having in mind the
construction of the thimble attached to the identity, one writes down
the hessian \\

\footnotesize 
\vskip-0.8cm
$$
\nabla^b_{m,\rhoh}\nabla^a_{n,\muh}S_G\l[U\r]\bigr|_{U=\mbb{I}}=
\frac{\beta}{2N}\delta^{ab}
{\l[2d\delta_{n,m}\delta_{\muh,\rhoh}-\delta_{n,m}+\delta_{n+\muh,m}+\delta_{n-\rhoh,m}-\delta_{n+\muh-\rhoh,m}-\delta_{\muh,\rhoh}\sum\limits_\nuh\l(\delta_{n+\nuh,m}+\delta_{n-\nuh,m}\r)\r]}
$$
\normalsize 
 It is easy to realize that the spectrum displays not only the 
$(V-1)(N_c^2-1)$ zero modes we discussed above, but also extra 
$d(N_c^2-1)$ ones\footnote{d is the dimension the theory lives in.}. This
does not come as a surprise: torons have shown up. There is an
extensive literature on the subject: the reader is referred to 
{\em e.g.} \cite{torons} and
references therein for an introduction to the results we will
mention in the following\footnote{We thank A. Ramos for having pointed
out ref \cite{torons} to us.}. 
Already for the simple case $d=2$ and $N_c=2$ one
gets a non-trivial case of study and since everything is
known in this setting, we have the chance to validate the results of 
a thimble formulation. 

Torons can be avoided by moving to a twisted action
$$
S_G\l[U\r]=\beta\sum_P f_P^{\l(t\r)}\l(U_P\r) \;\;\;\;\;\;\;\;\;
f_P^{\l(t\r)}\l(U_P\r) = 
\{f_P\l(z_{\muh\nuh}U_P\r) \;\mbox{for}\; P\in R_{\muh\nuh};
\;\; f_P\l(U_P\r) \;\mbox{for}\; P\notin R_{\muh\nuh} \}
$$
where $f_P\l(U_P\r)$ is the ordinary Wilson action density; 
$z_{\muh\nuh}=z_{\nuh\muh}^{-1}=\bar{z}_{\nuh\muh}=e^{2\pi i
  n_{\muh\nuh}/N}\in Z_N$ 
is the twist tensor and $R_{\muh\nuh}$ consists of a particular 
set of plaquettes. In our $d=2, N_c=2$ case $R_{\muh\nuh}$ simply reduces to the
plaquette which is named $P_0$ in figure 1. 
Minimum action configuration is now given by the so-called 
{\em twist-eater}. The construction of the latter starts with building
the {\em gauge tree} on which the links can be gauged to unity. 
\begin{figure}[ht] 
\centering
\includegraphics[scale=0.65]{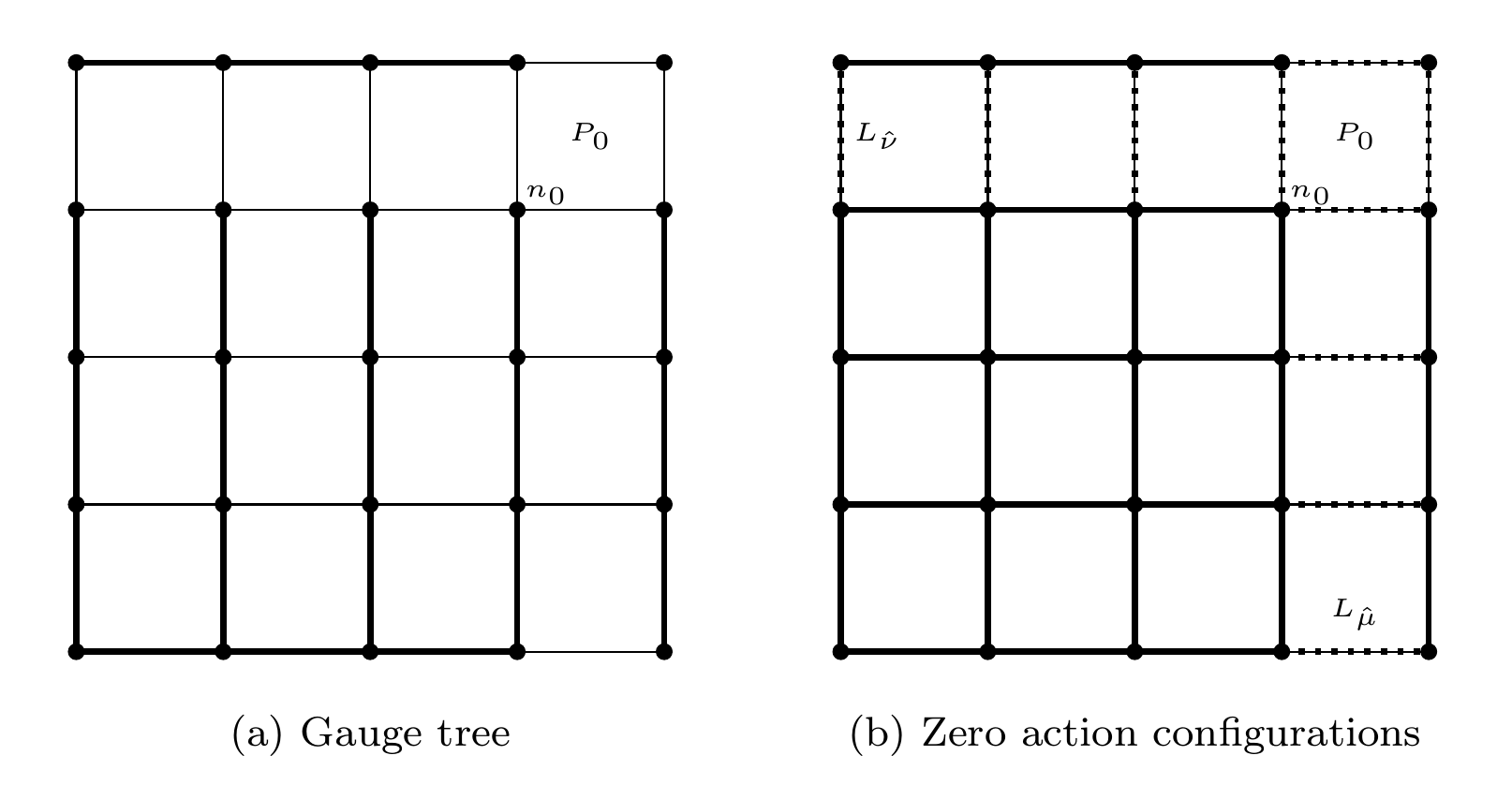}
\caption{The {\em gauge tree} construction for $d=2$. 
$P_0$ is the only plaquette which enters the action with a non-trivial
coefficient. $L_\muh$ and $L_\nuh$ are the ladders on which
links assume non-trivial values $G_\muh$ and $G_\nuh$ respectively.}
\label{fig:thimble_figure_stokes}
\end{figure}
On the two ladders $L_\muh$ and $L_\nuh$ links assume 
non-trivial values $G_\muh$ and $G_\nuh$ respectively. These values
are decided by the the twisted commutation relation
$G_\nuh G_\muh=z_{\muh\nuh}G_\muh G_\nuh$ which puts to zero the
contribution to the action coming from the plaquette $P_0$. 
It can be shown that the {\em twist-eater} is 
the global minimum, unaffected by torons. It is
thus the critical point whose thimble one should naturally 
start taking into account.

\section{First applications: SU(N) toy models}

While the previous section referred to a problem ($d=2$ $SU(2)$
Yang-Mills theory) for which we only have plans, we have already
worked out simple toy models like the $SU(N)$ one link models 
defined by
$$
S\l[U\r]=-\frac{\beta}{N}\mrm{Tr}\l(U\r) \;\;\;\;\;\;\;\;\;\;\;\;
Z\l(\beta\r)=\int\limits_{\mrm{SU}\l(N\r)}\mrm{d}U\,e^{\frac{\beta}{N}\mrm{Tr}\l(U\r)}=\sum\limits_{n=0}^\infty\frac{2!\cdots\l(N-1\r)!}{n!\cdots\l(n+N-1\r)!}\l(\frac{\beta}{N}\r)^{Nn}
$$
A natural observable is
$$
\langle\mrm{Tr}\l(U\r)\rangle=\frac{1}{Z}\int\limits_{\mrm{SU}\l(N\r)}\mrm{d}U\,\mrm{Tr}\l(U\r)\,e^{\frac{\beta}{N}\mrm{Tr}\l(U\r)}=N\frac{\de}{\de\beta}
\ln Z
$$
Also in this case the sign problem is artificial (one takes 
complex values for $\beta$). 
It is important to point out that for $SU(N)$ one finds 
exactly $N$ critical points (the elements of the center 
$Z_N$). All of them give a contribution to the correct computation 
of results, as {\em e.g.} depicted in figure 2 in the case of
$SU(3)$, where $|\beta|=5$ is kept fixed while 
$\arg \beta$ is varied.    

\begin{figure}[ht] 
\centering
\includegraphics[scale=0.65]{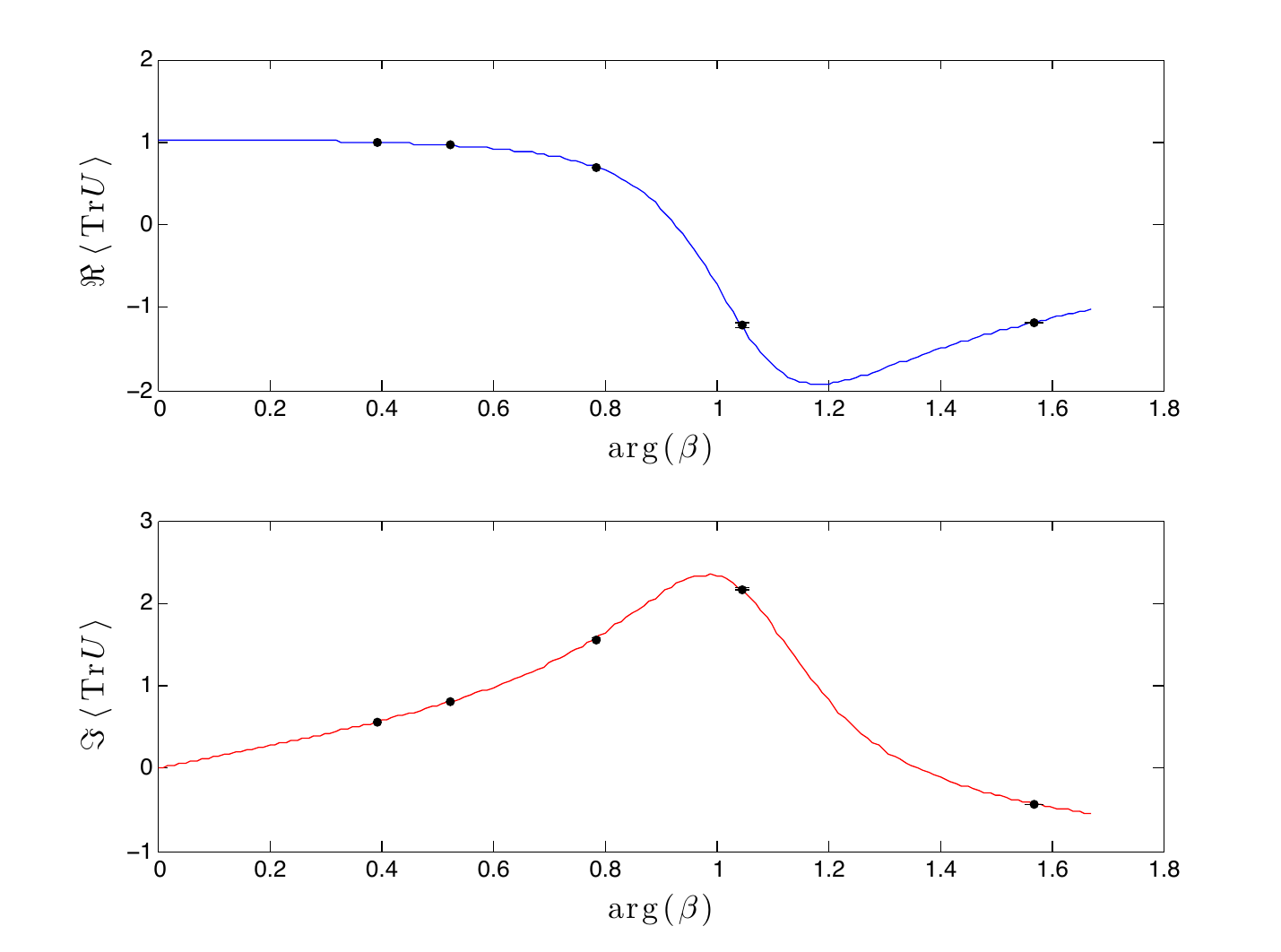}
\caption{Computation of $\langle\mrm{Tr}\l(U\r)\rangle$ at fixed 
$|\beta|=5$ for the $SU(3)$ one link model.}
\label{fig:thimble_figure_stokes}
\end{figure}

\section*{Conclusions}

Thimble regularization as a solution of the sign problem is still in
its infancy and there is quite a long way to go before 
we can have it working for Lattice QCD. Nevertheless, a formulation 
for gauge theories is there and simple toy models have already been 
successfully worked out.

\end{document}